\documentclass[11pt,a4paper]{article}
\usepackage{aas_macros,braket}
\usepackage{jcappub}
\usepackage[utf8]{inputenc}
\usepackage{amsmath}
\usepackage{amssymb}
\usepackage{bm}
\usepackage{url}
\usepackage[numbers,sort&compress]{natbib}
\usepackage[nottoc,notlot,notlof]{tocbibind}
\usepackage{wrapfig}
\usepackage[colorinlistoftodos]{todonotes}
\usepackage{color}
\usepackage{graphicx}
\usepackage{epsfig}
\usepackage{physics}
\usepackage[titletoc]{appendix}
\usepackage[normalem]{ulem}
\usepackage{framed}
\usepackage{tensor}
\usepackage{pdfcomment}
\usepackage{mathtools}
\usepackage{cancel}
\usepackage{geometry}
\usepackage{subcaption}
\usepackage{soul}
\usepackage{cleveref}
\usepackage{bm}
\crefrangeformat{equation}{(#3#1#4)-(#5#2#6)}

\usepackage{floatrow}
\newfloatcommand{capbtabbox}{table}[][\FBwidth]

\newcommand{\ba}{\begin{align}}
\newcommand{\ea}{\end{align}}

\newcommand{\pc}{\ensuremath{\mathcal{P}}}

\newcommand{\mpc}{\text{Mpc}}

\newcommand{\del}[2]{\frac{\partial #1}{\partial #2}}

\newcommand{\kpk}{\ensuremath{k_{\rm pk}}}
\newcommand{\TT}{\ensuremath{C_\ell^{TT}}}
\newcommand{\TE}{\ensuremath{C_\ell^{TE}}}
\newcommand{\EE}{\ensuremath{C_\ell^{EE}}}
\newcommand{\BB}{\ensuremath{C_\ell^{BB}}}
\newcommand{\rp}{\ensuremath{r_{\mathrm{pk}}}}
\newcommand{\lcdm}{\ensuremath{\Lambda{\text{CDM}}}}

\begin{document}
\begin{flushright}
\large \tt{CPPC-2022-09}
\end{flushright}
\title{Constraining primordial tensor features with the anisotropies of the Cosmic Microwave Background}
\author{Jan Hamann and Ameek Malhotra}
\affiliation{Sydney Consortium for Particle Physics and Cosmology, School of Physics, The University of New South Wales, Sydney NSW 2052, Australia}
\emailAdd{jan.hamann@unsw.edu.au}
\emailAdd{ameek.malhotra@unsw.edu.au}

\abstract{
It is commonly assumed that the stochastic background of gravitational waves on cosmological scales follows an almost scale-independent power spectrum, as generically predicted by the inflationary paradigm. However, it is not inconceivable that the spectrum could have strongly scale-dependent features, generated, e.g., via transient dynamics of spectator axion-gauge fields during inflation. Using the temperature and polarisation maps from the \textit{Planck} and BICEP/Keck datasets, we search for such features, taking the example of a log-normal bump in the primordial tensor spectrum at CMB scales.  We do not find any evidence for the existence of bump-like tensor features at present, but demonstrate that future CMB experiments such as LiteBIRD and CMB-S4 will greatly improve our prospects of determining the amplitude, location and width of such a bump.  We also highlight the role of delensing in constraining these features at angular scales $\ell\gtrsim 100$.
}

\maketitle
\flushbottom

\section{Introduction}
The inflationary paradigm~\cite{Guth:1980zm,Sato:1980yn,Linde:1981mu,Albrecht:1982wi,Linde:1983gd} is not only a compelling solution to the horizon and flatness problems of hot big bang cosmology, but also provides a means to naturally generate the seeds of the observed Cosmic Microwave Background (CMB) anisotropies and the Large Scale Structure (LSS)~\cite{Mukhanov:1981xt,Hawking:1982hg,Guth:1982ec,Starobinsky:1982ee,Bardeen:1983qw} via quantum fluctuations. 
Cosmological observations are consistent with a power spectrum of scalar (density) fluctuations that is adiabatic, Gaussian and nearly (but not exactly) scale-invariant, in excellent agreement with the predictions of the simplest single field slow roll (SFSR) models~\cite{Akrami:2018odb}. 
Another universal prediction of inflation is the existence of a stochastic gravitational wave background produced during this epoch~\cite{Grishchuk:1974ny,Starobinski:1979JETPL,Rubakov:1982df,Fabbri:1983us}. 

Although such primordial gravitational waves (PGW) have not yet been detected, their effects could show up in a wide variety of cosmological observables. Most notably, PGW contribute to both the temperature and polarisation anisotropies of the CMB~\cite{Fabbri:1983us,Sachs:1967sw,Abbott:1984gw,Starobinskii:1985gw,Kamionkowski:1996zd,Seljak:1996gy,Zaldarriaga:1996xe,Kamionkowski:1996ks}, and this fact can be used to constrain their amplitude~\cite{BK18,Tristram:2021tvh,Galloni:2022mok,Paoletti:2022anb,Namikawa:2019tax,Clarke:2020bil,Ng:2021waj}. Indeed, the most stringent bounds on the amplitude of these PGW come from the CMB which constrains the tensor-scalar ratio to $r < 0.032$~\cite{Tristram:2021tvh} at $95\%$ CL for a nearly scale-invariant power spectrum of tensor fluctuations, as expected in SFSR models.

Interferometric detection with next generation detectors like ET~\cite{Punturo:2010zz} and LISA~\cite{amaroseoane2017laser} may also be a possibility, but only for inflationary models departing strongly from SFSR dynamics on direct detection scales~\cite{Maggiore:2019uih,LISACosmologyWorkingGroup:2022jok}. Additionally, PGW could also be detected via Pulsar Timing arrays, through their imprints on large scale structure (`tensor fossils'), spectral distortions of the CMB, as well as through the gravitational lensing effects of PGW, see~\cite{Guzzetti:2016mkm} for an overview.

The $B$-mode polarisation of the CMB remains the most promising avenue to detect these primordial tensor perturbations, keeping in the mind the projected sensitivities as well the foreground/noise sources affecting the various probes mentioned above. A detection of these primordial tensor perturbations would be extremely significant since their amplitude in SFSR models is directly related to the energy scale of inflation $V_\mathrm{inf} \simeq 3\pi^2 A_\mathrm{s}M_\mathrm{pl}^4\,r/2$~\cite{Guzzetti:2016mkm} and would allow a precise reconstruction of the inflaton potential in the observable window~\cite{Copeland:1993zn,Peiris:2006ug,Hamann:2008pb}. Interestingly, certain well-motivated single field inflationary models are currently in excellent agreement with the CMB data~\cite{Akrami:2018odb}, e.g.~the Starobinsky model~\cite{Starobinsky:1980te,Starobinsky:1983zz,Kehagias:2013mya}, and predict values of $r$ within the reach of next generation of CMB experiments. For these reasons, it is easy to understand why the search for PGWs is  an important science goal for future probes like the BICEP array~\cite{Moncelsi:2020ppj}, Simons Observatory~\cite{SimonsObservatory:2018koc}, CMB-S4~\cite{CMB-S4:2016ple} and LiteBIRD~\cite{LiteBIRD:2022cnt}.

While a detection of PGW in itself would be extremely valuable, additional information on the production mechanism could be gleaned from measuring the shape of the GW spectrum. For PGW generated from vacuum fluctuations, the shape of the spectrum is a power law with spectral index related to the tensor-scalar ratio as $n_\mathrm{t} \simeq -r/8$. This  `tensor consistency relation' is valid for PGW arising from SFSR dynamics~\cite{Copeland:1993ie} and in the event of a $B$-mode detection, provides a way to confirm whether the observed PGW are arising from vacuum fluctuations or not. Unfortunately however, testing this relation appears out of reach with CMB data alone, even with the sensitivity of CMB-S4~\cite{CMB-S4:2016ple} or LiteBIRD~\cite{Campeti:2020xwn,Paoletti:2022anb}. On the other hand, large deviations from the consistency relation could still be observed with these experiments. Such deviations would signal a departure from SFSR dynamics at CMB scales which is possible in inflationary models with additional fields capable of sourcing PGW, e.g.~\cite{Cook:2011hg,Barnaby:2011qe,Barnaby:2012xt,Mukohyama:2014gba,Maleknejad:2016qjz,Dimastrogiovanni:2016fuu,Garcia-Bellido:2016dkw,Thorne:2017jft,Domcke:2018eki,Campeti:2020xwn,Fujita:2022jkc,Campeti:2022acx}.

In this paper we study one such example of a deviation from a power-law spectrum of tensor perturbations, namely a bump-like PGW feature at CMB scales. Such a feature is typical of GW sourced from spectator axion-gauge fields during inflation, leading to a strong scale-dependence. Parameterising the spectrum of these sourced PGW as a log-normal function, we present constraints on the amplitude, width and location of the peak of the log-normal using temperature and polarisation data from the \textit{Planck} and BICEP/Keck datasets and forecast the discovery potential of CMB-S4 and LiteBIRD for this scenario.

The paper is organised as follows: in Section~\ref{sec:Tensor_spectrum} we describe the tensor power spectrum parameterisation and discuss possible inflationary models where such a spectrum may arise. We also describe the effects of such a tensor power spectrum shape on the CMB temperature and polarisation anisotropies. 
In Section~\ref{sec:constraints} we present constraints on the model parameters from current data and forecast sensitivities of future CMB experiments.
Finally, we conclude in Section~\ref{sec:conclusion}.

\section{Tensor modes from Inflation}
\label{sec:Tensor_spectrum}
In this section we first present the tensor power spectrum parameterisation used to describe the bump-like feature. We then discuss the effect of such a tensor spectrum shape on the CMB temperature and polarisation anisotropies and conclude this section with a discussion of inflationary models where one can expect such power spectrum shapes.

\subsection{Log-normal spectrum}
\begin{figure}
    \begin{subfigure}{0.45\textwidth}
   \includegraphics[width=\linewidth]{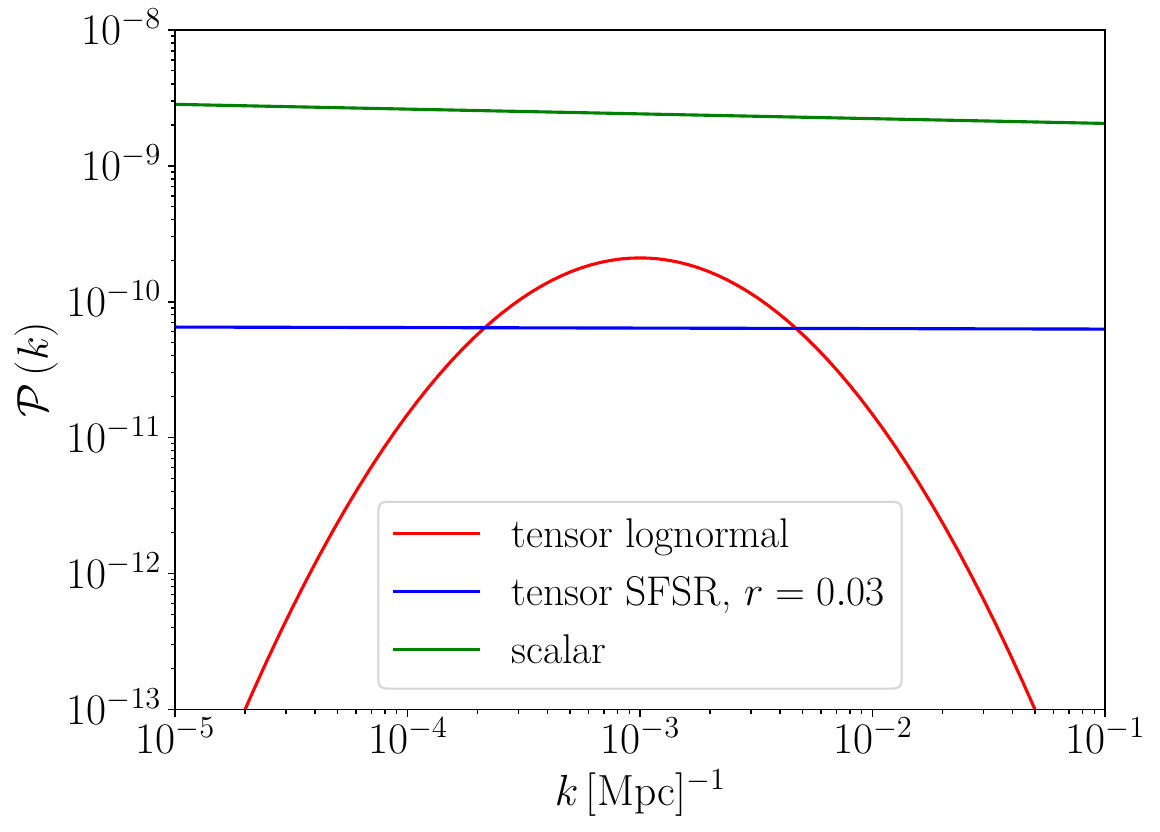}
   \end{subfigure}
   \begin{subfigure}{0.45\textwidth}
      \includegraphics[width=\linewidth]{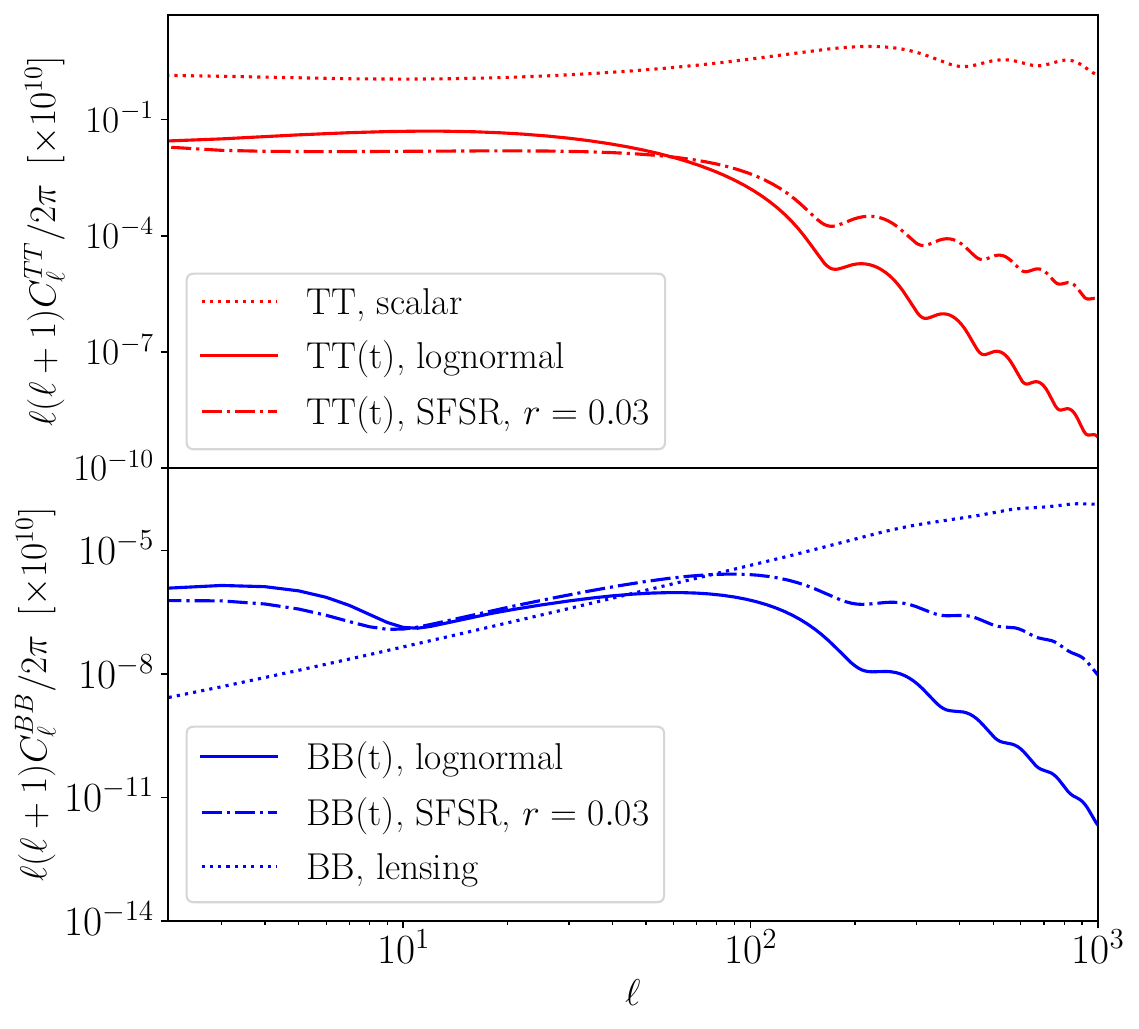}
   \end{subfigure}
    \caption{Left : Power spectra with tensor spectrum parameters $\rp=0.1,~\sigma=1$ and $k_{\rm pk}=10^{-3}$. Right: The temperature and $B$-mode anisotropies for the same value of the tensor power spectrum parameters. }
    \label{fig:power_spec_cl}
\end{figure}

We parameterise the tensor power spectrum on CMB scales as a log-normal with
\begin{align}
    \label{eq:pt_lognormal}
    \pc_\mathrm{t}(k) = \rp A_\mathrm{s} \exp\left[-\frac{\left(\ln \left(k/k_{\rm pk}\right)\right)^2}{2\sigma^2}\right].
\end{align}
Here $\sigma$ denotes the width of the log-normal, $\kpk$ the location of the peak and $\rp$ the rescaled amplitude of the tensor power spectrum at the peak scale in terms of the scalar amplitude $A_\mathrm{s}=2.09\times 10^{-9}$ at the pivot scale $k_\mathrm{p}=0.05\,\mpc^{-1}$.  When $\sigma\gg 1$, the power spectrum becomes degenerate with a flat spectrum on scales
\begin{align}
    -\sigma\lesssim\log k/\kpk\lesssim\sigma.
\end{align}
For comparison, the power spectrum of tensor perturbations from vacuum fluctuations during inflation has the following form,
\begin{align}
    \label{eq:pt_sfsr}
    \pc_\mathrm{t}^{\rm vac}(k) = r A_\mathrm{s} \left(\frac{k}{k_\mathrm{p}}\right)^{n_\mathrm{t}}
\end{align}
where $r$ denotes the vacuum tensor to scalar ratio and $n_\mathrm{t}$ the tensor spectral tilt. The two power spectra are plotted in the left panel of Figure~\ref{fig:power_spec_cl} taking $r=0.03$ and $n_\mathrm{t}$ given by the consistency relation for the SFSR spectrum and $\rp=0.1,\;\sigma=1$ and $\kpk=10^{-3}$ for the log-normal.

\subsubsection{CMB anisotropies}
\label{sec:tensors_cmb}
Much like the scalar perturbations, tensors also contribute to the CMB anisotropies and this contribution can be expressed as
\begin{align} 
    C_{\ell,\rm t}^{XY} = 4\pi\int d\ln k\,\pc_\mathrm{t}(k) \,\Delta_{\ell,\rm t}^{X}(k)\Delta_{\ell,\rm t}^{Y}(k).
\end{align}
Here $X,Y = T,E,B$ and $\Delta_{\ell,\rm t}^X$ denotes the corresponding transfer function for the tensor source. The analytical expressions for these transfer functions can be found in~\cite{Kamionkowski:1996zd,Seljak:1996gy,Zaldarriaga:1996xe,Kamionkowski:1996ks}, and they can be numerically computed using Boltzmann codes such as \texttt{CLASS}~\cite{Blas:2011rf} or \texttt{CAMB}~\cite{Lewis:1999bs}. 

Physically, tensor modes generate a quadrupolar anisotropy in the radiation density field at last scattering and reionisation, and thus tensor perturbations lead to anisotropies in both temperature and polarisation of the CMB. Note that for non-chiral primordial tensor spectra, the correlations $\langle TB \rangle$ and  $\langle EB \rangle$ vanish due to the fact that $B$-modes are parity-odd whereas $T$ and $E$ are not~\cite{Lue:1998mq}. 

In general, the tensor contribution is relevant mainly on the largest angular scales since tensor modes decay rapidly on sub-horizon scales ($\ell\gtrsim 100$ at recombination). For the temperature anisotropies, the tensor contribution is smaller than the scalar one by a factor of the tensor-scalar ratio $r$ on these large scales. As for the $B$-mode polarisation, for $r\gtrsim 0.001$ this is dominated by the primordial tensor perturbations on these scales since scalars cannot source $B$-modes at linear order. 

However, scalars do generate $B$-mode polarisation at second order via the lensing of the $E$-modes~\cite{Zaldarriaga:1998ar} and these lensing-induced $B$-modes dominate the primordial $B$-mode spectrum at smaller scales ($\ell\gtrsim 100$). In the $B$-mode polarisation angular power spectrum one can also see the reionization bump ($\ell\leq 10$)~\cite{Zaldarriaga:1996xe} and the recombination bump ($\ell\sim 80$)~\cite{Zaldarriaga:1996ke}, corresponding to scales that re-enter the horizon at those times. 

We plot the tensor contribution to the temperature and $B$-mode angular power spectra in the right panel of Figure~\ref{fig:power_spec_cl}. The difference between the SFSR prediction and the log-normal spectrum can also be understood from the same figure. The additional power on scales close to the peak scale $k=10^{-3}$ leads to an enhancement of the temperature as well as the polarisation anisotropies on large angular scales. Away from the peak scales, the anistropy power spectrum in the log-normal case falls off rapidly relative to the SFSR one.

\begin{figure}
    \includegraphics[width=0.96\linewidth]{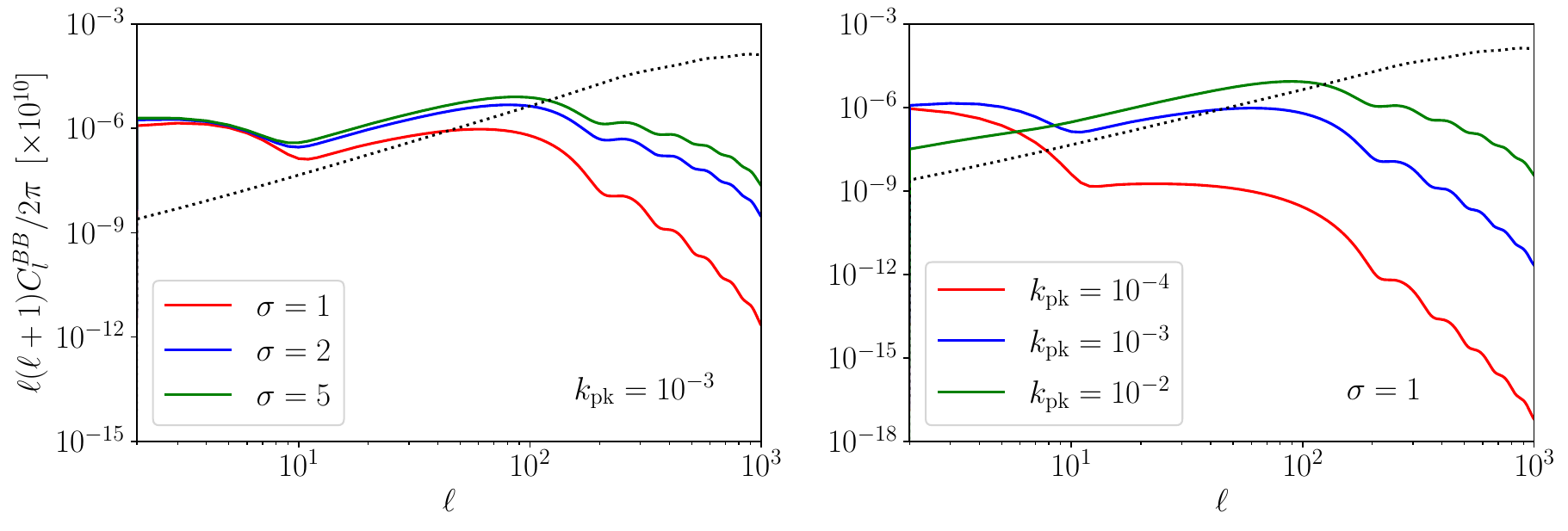}
    \caption{Effect of varying $\sigma$ and $\kpk$ on $C_\ell^{BB}$ taking $\rp=0.1$. The black dotted lines show the $B$-modes from lensing.}
    \label{fig:bb_sigma_kpk}
\end{figure}

The effect of varying the log-normal parameters on the $B$-mode spectrum is illustrated in Figure~\ref{fig:bb_sigma_kpk}. The location of the peak scale $\kpk$ can be directly related to the angular scale at which the anisotropies are enhanced, whereas the peak width $\sigma$ controls the range of scales around the peak where this happens. Roughly speaking, peaks at $\kpk=10^{-4},10^{-3},10^{-2}$ can be mapped to an enhancement of power centered around angular scales of $\ell\sim 2,\,10\;\text{and}\;100$ respectively. 

\subsubsection{Inflationary Models}
A peaked primordial tensor power spectrum is characteristic of transient phenomena occurring during inflation that source gravitational waves on scales of order the horizon size at that time. The prototypical examples of this type are inflationary models with a spectator sector involving an axion coupled to a gauge field~\cite{Mukohyama:2014gba,Maleknejad:2016qjz,Dimastrogiovanni:2016fuu,Garcia-Bellido:2016dkw,Thorne:2017jft,Domcke:2018eki,Campeti:2020xwn,Fujita:2022jkc,Campeti:2022acx}. The spectator sector Lagrangian in this case can be written as
\begin{align}
    \label{eq:lagrangian}
    \mathcal{L}_\mathrm{spec} = -\frac{1}{2}(\partial\chi)^2 -U(\chi) - \frac{1}{4}F_{\mu\nu} \tilde{F}^{\mu,\nu} +\frac{\lambda\,\chi}{f} F_{\mu\nu} \tilde{F}^{\mu,\nu}  
\end{align}
where $\chi$ represents the axion-like field, $U(\chi)$ is the axion potential, $f$ the decay constant, $F_{\mu\nu}$ the field strength tensor of the gauge field and $\lambda$ a coupling constant. The specific shape of the GW spectrum in such models arises from the fact that one of the gauge field helicities experiences a transient instability at horizon crossing and gets enhanced relative to the other. Thus, in general the GW spectrum sourced in these models is chiral, i.e. $\pc_R \neq \pc_L$. Although one can test for chirality through the observation of non-zero $\langle TE \rangle$ or $\langle EB \rangle$ cross-spectra~\cite{Lue:1998mq,Sorbo:2011rz,Thorne:2017jft}, in our analysis we only concern ourselves with the overall shape and amplitude of the tensor power spectrum and neglect these parity violating correlations in obtaining the constraints in Section~\ref{sec:constraints}. In general the signal is much weaker in the cross-spectra than in the corresponding $\langle B B \rangle$ spectrum, making them harder to detect \cite{Thorne:2017jft}.

\section{Constraints on model parameters \label{sec:constraints}}

Constraints on the axion gauge-field model parameters of equation~\eqref{eq:lagrangian}, specifically upper bounds on the effective coupling between the axion and the gauge field were also obtained in~\cite{Campeti:2022acx} for two different models of the axion potential $U(\chi)$. The analysis of~\cite{Campeti:2022acx} was carried out for fixed values of the peak width and location using the profile likelihood approach. However, our analysis cannot be directly compared with theirs since varying the effective coupling also affects the sourced scalar spectra whereas here we work at the level of the log-normal parameters which affects only the primordial tensor power spectrum through equation~\eqref{eq:pt_lognormal}.

\subsection{Constraints from \textit{Planck} + BK18}
\label{sec:pl_bk18}
For the analysis of our model comprising the base $\Lambda$CDM parameters plus our additional tensor power spectrum parameters $\{\rp,\kpk,\sigma\}$, we use the following data:
Firstly, the \textit{Planck} low-$\ell$ temperature+polarisation and high-$\ell$ $TTTEEE$ likelihood~\cite{Aghanim:2019ame} and the \textit{Planck} lensing likelihood~\cite{Aghanim:2018oex}. We shall refer to this combination as \textit{Planck} hereafter. 
Secondly, we also use the most recent BICEP/Keck data release (BK18 hereafter) which contains polarisation data in the multipole range $20<\ell<330$ obtained from the BICEP2, Keck Array and BICEP3 experiments~\cite{BK18}. 

We impose flat priors on the base $\Lambda$CDM parameters as well as flat priors on $\rp,\sigma$ and $\log_{10}\kpk$ in the ranges shown in Table~\ref{tab:prior}.  The CMB power spectra are computed with \texttt{CAMB}~\cite{Lewis:1999bs} and the parameter space is explored using a Markov Chain Monte Carlo sampler~\cite{Lewis:2002ah,Lewis:2013hha}, through its interface with \texttt{Cobaya}~\cite{Torrado:2020dgo}. The resulting chains are analysed with \texttt{GetDist}~\cite{Lewis:2019xzd}. 

Note that we assume in our analysis that the scale-independent vacuum contribution to the tensor power spectrum has a much lower amplitude than the log-normal sourced one, so we do not include the variation of $r$ and instead set it to zero. Current data do not indicate the presence of such vacuum tensor perturbations and for values of $r$ much smaller than $\rp$ the constraints on $\rp$ will not be significantly affected. We will revisit this assumption when doing the Fisher forecasts in Section~\ref{sec:forecasts}.

\begin{figure}
    \centering
    \includegraphics[width=0.75\linewidth]{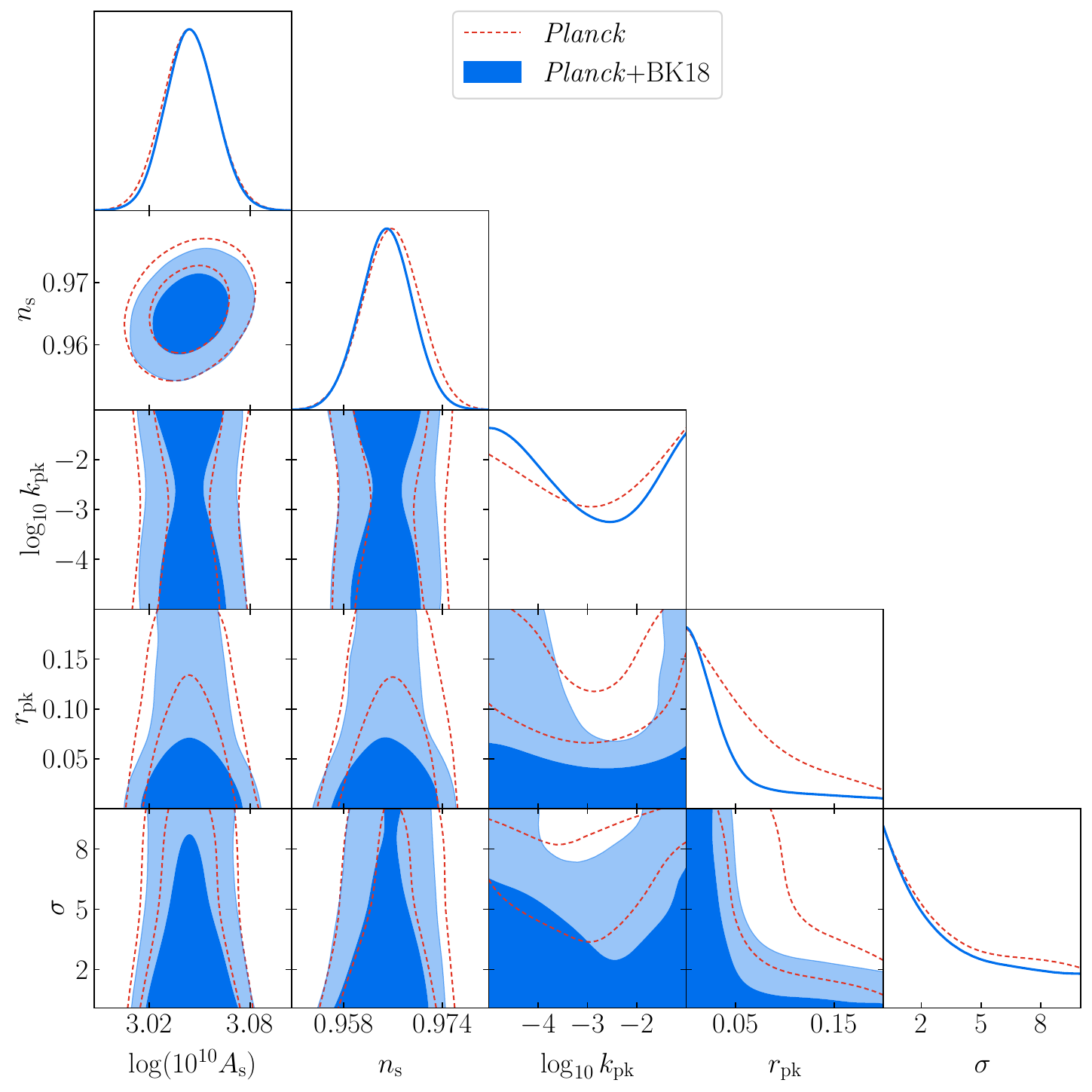}
    \caption{Marginalised $68\%$ and $95\%$ contours for the scalar and tensor power spectrum parameters. For comparison we also show the same contours obtained using only the \textit{Planck} data, this mainly relaxes the constraints on $\rp$.}
    \label{fig:triangle_ps_pars}
\end{figure}

\begin{table}
    \centering
    \renewcommand\arraystretch{1.1}
    \begin{tabular}{c c}
%\hline
Parameter & range\\
\hline
  \boldmath{$\rp$}   &  $[0,1/5]$\\
\boldmath{$\sigma $} & $[1/10,10]$\\
\boldmath{$\log_{10}\kpk$} & $[-5,-1]$\\
%\hline
\end{tabular}
    \caption{Prior ranges for the log-normal tensor power spectrum parameters. }
    \label{tab:prior}
\end{table}
The joint posterior distributions for the scalar and tensor power spectrum parameters are shown in Figure~\ref{fig:triangle_ps_pars} (for posterior contours of all 9 parameters we refer the reader to Figure~\ref{fig:triangle_all} in the Appendix). The tensor spectrum parameters are found to be mostly uncorrelated with the scalar spectrum ones. We also see that the strongest constraints on $r$ come from the region $10^{-3}<\kpk<10^{-2}$ which is not surprising since features at these scales mainly affect anisotropies in the multipole range $20\lesssim\ell\lesssim 300$, i.e., exactly the range covered by the BK18 $B$-mode data. 

\begin{figure}
    \centering
    \includegraphics[width=0.5\linewidth]{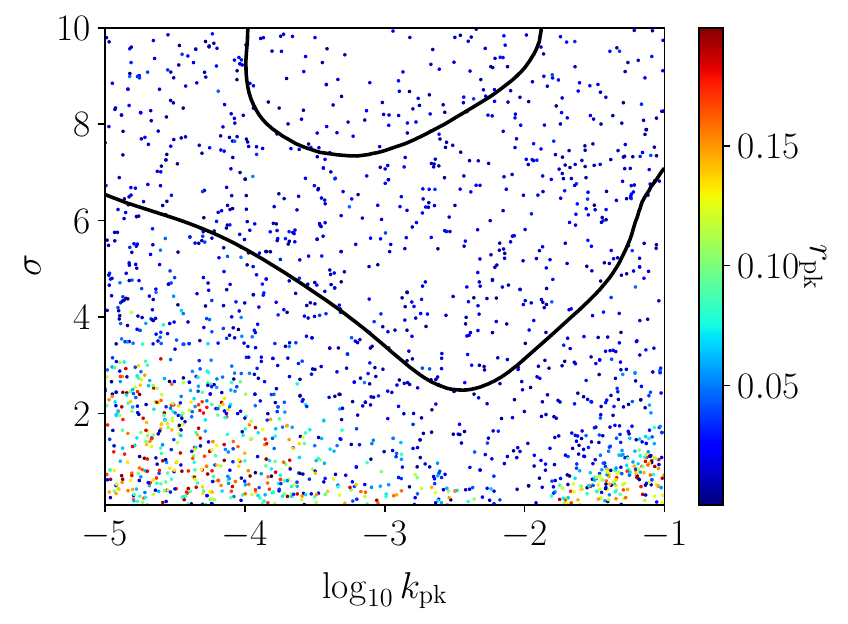}
    \caption{Distribution for the tensor power spectrum parameters using \textit{Planck}+BK18 data.}
    \label{fig:3d_r}
\end{figure}

For large $\sigma$ the constraints on $r$ reduce to the flat/power-law constraints, as expected. Figure~\ref{fig:3d_r} also shows that large values of $r \gtrsim 0.1$ are only allowed in the region $\kpk\lesssim 10^{-4}$ or $\kpk\gtrsim 10^{-2}$ and for $\sigma \lesssim 2$.  The parameter limits are presented in Table~\ref{tab:par_limits}. The best fit point is found to be $\{\rp=0.04,\,\sigma=0.43,\,\kpk=2\times 10^{-2}\}$ with a $\Delta\chi^2$ relative to base $\lcdm$ of $\Delta\chi^2=-0.22$.  This result is fully compatible with the absence of tensor modes in the \textit{Planck}+BK18 data.

\begin{table}
    \centering
    \begin{tabular} { l  c}

 Parameter & 68\% limit \\ 
\hline
{\boldmath $\sigma$} & $<4.83$ \\  

{\boldmath $\rp$} & $< 0.0460$  \\
\end{tabular}
\caption{Parameter limits for the tensor log-normal parameters from \textit{Planck}+BK18 data. {Note that these upper limits are prior dependent and may change considerably for different choices of the prior ranges for the three tensor power spectrum parameters.}}
\label{tab:par_limits}
\end{table}

\subsection{Forecasts with LiteBIRD + CMB-S4 \label{sec:forecasts}}
In this section we present forecasts of the ability of next generation CMB experiments to constrain the log-normal tensor parameters. We first take the example of LiteBIRD which is a proposed space-based CMB experiment led by JAXA with the goal of mapping the temperature and polarisation anisotropies of the CMB in the multipole range $2\leq\ell\leq200$~\cite{LiteBIRD:2022cnt}. Thus, the reionization bump ($\ell\lesssim 10$) as well as the recombination bump ($\ell\sim 80$) will both be accessible to LiteBIRD. Note that a similar analysis of the detectability of such signals with LiteBIRD was also performed in Ref.~\cite{Thorne:2017jft}, but only for values of $\sigma$ in the range $2\leq\sigma\leq 10$ and taking $\kpk=7\times 10^{-5},\,5\times10^{-3}\mpc^{-1}$.

To estimate the constraining power of LiteBIRD, we perform a simple Fisher matrix forecast\footnote{The results presented in this paper are based on only varying the tensor parameters in the Fisher forecast. We checked that including the $\lcdm$ parameters does not significantly affect the estimates for the tensor parameters' expected uncertainties.} to evaluate the detection prospects of the log-normal tensor spectrum. The Fisher matrix for the parameters $\vec{\theta} = (r_\mathrm{peak},\sigma,\kpk)$ can be written as~\cite{Wu:2014hta},
\begin{align}
    F_{ij} = f_{\rm sky}\sum_{\ell=2}^{\ell_{\rm max}}\frac{2\ell+1}{2}\Tr\left[\mathbf{C}_\ell^{-1}\del{\mathbf{C}_\ell}{\theta_i}\mathbf{C}_\ell^{-1}\del{\mathbf{C}_\ell}{\theta_j}\right],
\end{align}
with the marginalised $1\sigma$ error on the parameter $\theta_i$  given by
\begin{align}
    \sigma_i = \sqrt{\left(F^{-1}\right)_{ii}}.
\end{align}
The matrix $\mathbf{C}_\ell$ is defined as,
\begin{align}
    \begin{pmatrix}
    \TT +N_\ell^{TT} & \TE & 0 \\
    \TE & \EE+N_\ell^{EE} & 0 \\
    0 & 0 & \BB + N_\ell^{BB}
    \end{pmatrix}
\end{align}
We also have
\begin{align}
    \frac{1}{N_\ell^{X,\mathrm{inst}}} = \sum_{\nu_i}\frac{1}{N_{\ell,\nu_i}^X}
\end{align}
where the instrument noise spectra at an observing frequency $\nu$ can be written as,
\begin{align}
    N_{\ell,\nu}^{X} = \Delta^X_\nu \exp\left[\ell(\ell+1)\frac{\theta^2_{\rm FWHM}}{8\log 2}\right]
\end{align}
with $X=TT,EE,BB$.  We take $f_{\rm sky}=0.65,\,\ell_{\rm max}=200$ and adopt the LiteBIRD instrument noise specifications given in Eq.~(3.1) of Ref.~\cite{Hazra:2018eib} for temperature. For the $B$-mode polarisation we also include foregrounds in addition to the instrument noise and then use the residual foregrounds plus post component separation noise as described in~\cite{Campeti:2020xwn}.\footnote{The corresponding noise data were made available by the authors of Ref.~\cite{Campeti:2020xwn} at \url{https://github.com/pcampeti/SGWBProbe}.} The lensing $B$-modes also act as a noise component in the search for the primordial signal and are included in the calculation of the Fisher matrices.  

Much like the instrument noise, the presence of foregrounds and the lensing $B$-modes also hinders our ability to cleanly detect the primordial $B$-mode signal. Separating the contributions of these foregrounds and lensing $B$-modes is crucial for the detection of the inflationary gravitational wave background . Synchrotron and thermal dust emission from diffuse sources in the galaxy constitute the two main types of polarised foregrounds on large angular scales. Typically, one utilises the fact that the frequency dependence of these foregrounds is different from the CMB signal to separate their contributions to the observed polarisation and temperature maps~\cite{Dickinson:2016xyz}. Thus, having a large frequency coverage is vital to accurately constrain the primordial $B$-modes.

Removing the lensing contaminant instead requires high-resolution maps of the $E$-mode polarisation as well as the CMB lensing potential~\cite{Kesden:2002ku,Hirata:2003ka} (or even the cosmic infrared background~\cite{Planck:2013qqi}). These are used to first estimate the lensing $B$-modes and then this estimate is used to delens the $B$-mode signal. The feasibility of this procedure in reducing the lensing $B$-mode power (and also in acoustic peak sharpening) has already been demonstrated with \textit{Planck} \cite{Aghanim:2018oex} and will be greatly improved with next generation experiments like CMB-S4~\cite{CMB-S4:2016ple,CMB-S4:2020lpa}.

\begin{figure}
    \centering
    \includegraphics[width=0.48\linewidth]{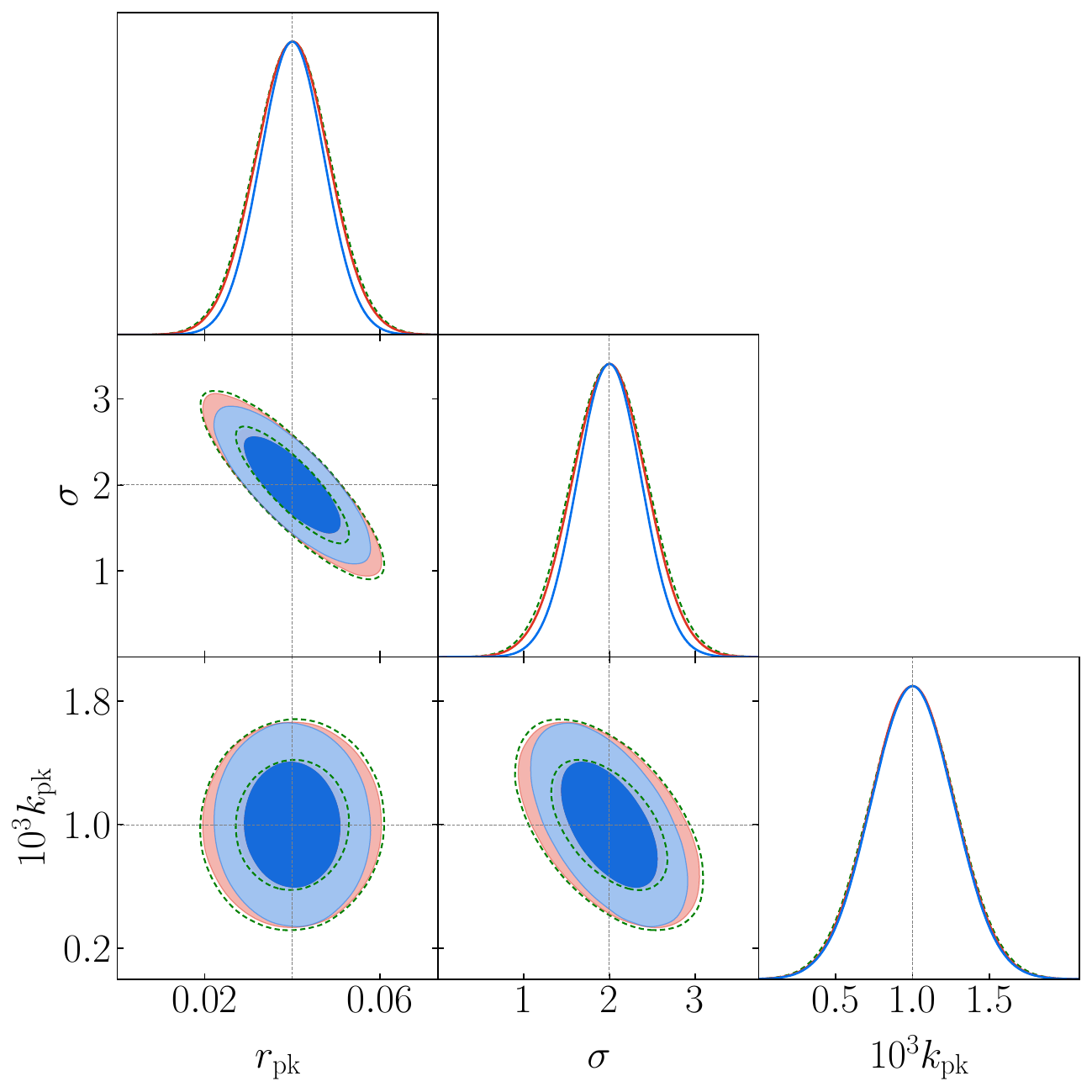}
    \includegraphics[width=0.5\linewidth]{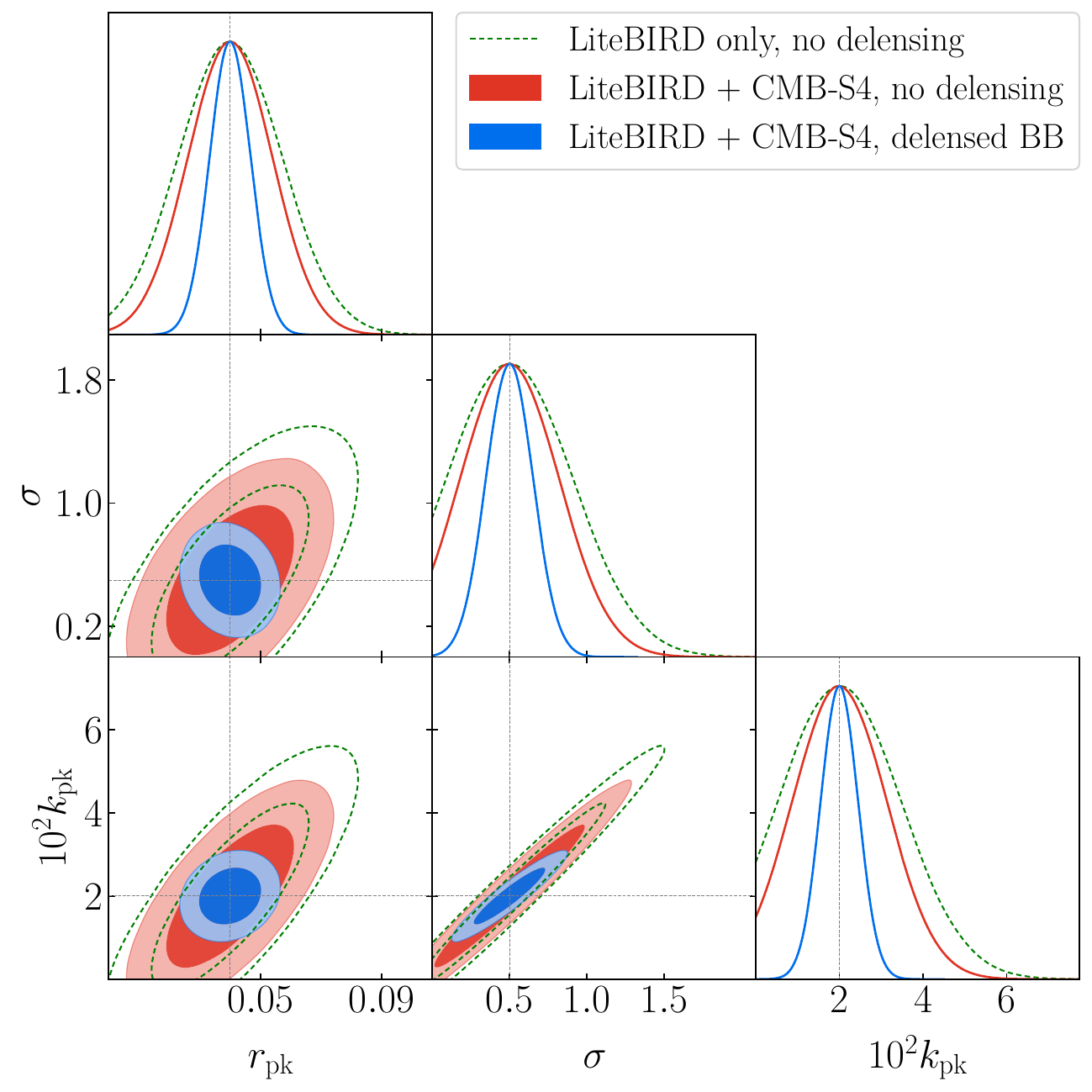}
    \caption{Fisher forecast for the tensor parameters assuming fiducial models with tensor features that have a chance of being detected by LiteBIRD and CMB-S4, taking $\rp=0.05,\,\sigma=2,\,k_{\rm pk}=10^{-3}$ (left) and $\rp=0.04,\,\sigma=0.5,\,k_{\rm pk}=2\times 10^{-2}$ (right). The blue ellipses show the estimates in the case where the CMB-S4 data are delensed with $A_\mathrm{delens}=0.8$.
    \label{fig:Fisher1}}
\end{figure}

In the left panel of Figure~\ref{fig:Fisher1} we present the forecast for the parameter set \mbox{$\{\rp=0.04$}, $\sigma=2,\,\kpk=10^{-3}\}$. Although this parameter set is quite close to the upper limits from the \textit{Planck}+BK18 data, it is useful to understand the constraining ability of LiteBIRD. 

One can see that the tightest constraints can be achieved for $\rp$ whereas precise measurements of $\sigma$ require a significant amount of delensing. This should be clear from the fact that to measure the width of the primordial spectrum accurately, we need to detect the primordial $B$-mode signal over a larger range of scales. However, significant delensing using CMB data is not possible with LiteBIRD alone since it requires information at small scales which are inaccessible to the experiment. This issue could however be overcome by using future external datasets such as those from CMB-S4. 

In the right panel of Figure~\ref{fig:Fisher1}, we also present forecasts for the case of $\{\rp=0.04,\,\sigma=0.5,\,\kpk=2\times 10^{-2}\}$, corresponding to a parameter set which produces features on scales $\ell\sim 100$. In this case, using LiteBIRD data alone, the constraints on $\rp$ are weaker since this corresponds to scales  where the $B$-mode signal is lensing dominated. Significant delensing needs to be achieved to precisely constrain this scenario. For this purpose, we consider the possibility of delensing using the high-resolution capabilities of CMB-S4~\cite{CMB-S4:2016ple,CMB-S4:2020lpa}.

CMB-S4 forecasts for $r$ typically assume $f_\mathrm{sky}=0.03$ and we take the same value here for the Fisher forecasts of the lognormal parameters. This specific choice of $f_\mathrm{sky}$ arises from an optimisation procedure which takes into account delensing requirements, foreground mitigation and reproducibility of results across the sky \cite{CMB-S4:2020lpa}. We also take a delensing fraction $A_\mathrm{delens}=0.8$ which represents an $80\%$ reduction in the lensing $B$-mode power spectra. The analysis of~\cite{CMB-S4:2020lpa} suggests that achieving this value is not entirely unrealistic. The CMB-S4 multipole range for the delensed $B$-mode data is taken to be $30<\ell<300$ and the instrument noise specifications are those corresponding to the `SAT' configuration and are available at this link\footnote{\url{https://cmb-s4.uchicago.edu/wiki/index.php/Delensing_sensitivity_-_updated_sensitivities,_beams,_TT_noise}}. We do not take into account foreground residuals for CMB-S4 and assume that these have been perfectly subtracted from the data. In reality, imperfect subtraction will lead to the presence of foreground residuals which will again act as an additional noise towards the detection of the primordial signal but this will not degrade sensitivity to the tensor-scalar ratio by more than $10^{-3}$ \cite{CMB-S4:2020lpa}, which is smaller than the $B$-mode amplitudes in our fiducial models.  The resulting estimates for this setup are shown in the same figure. We can see that a significant increase in constraining power is obtained with the addition of CMB-S4 and a delensing level $A_\mathrm{delens}=0.8$.

We conclude this section with two more examples. \begin{figure}
    \centering
    \includegraphics[width=0.46\linewidth]{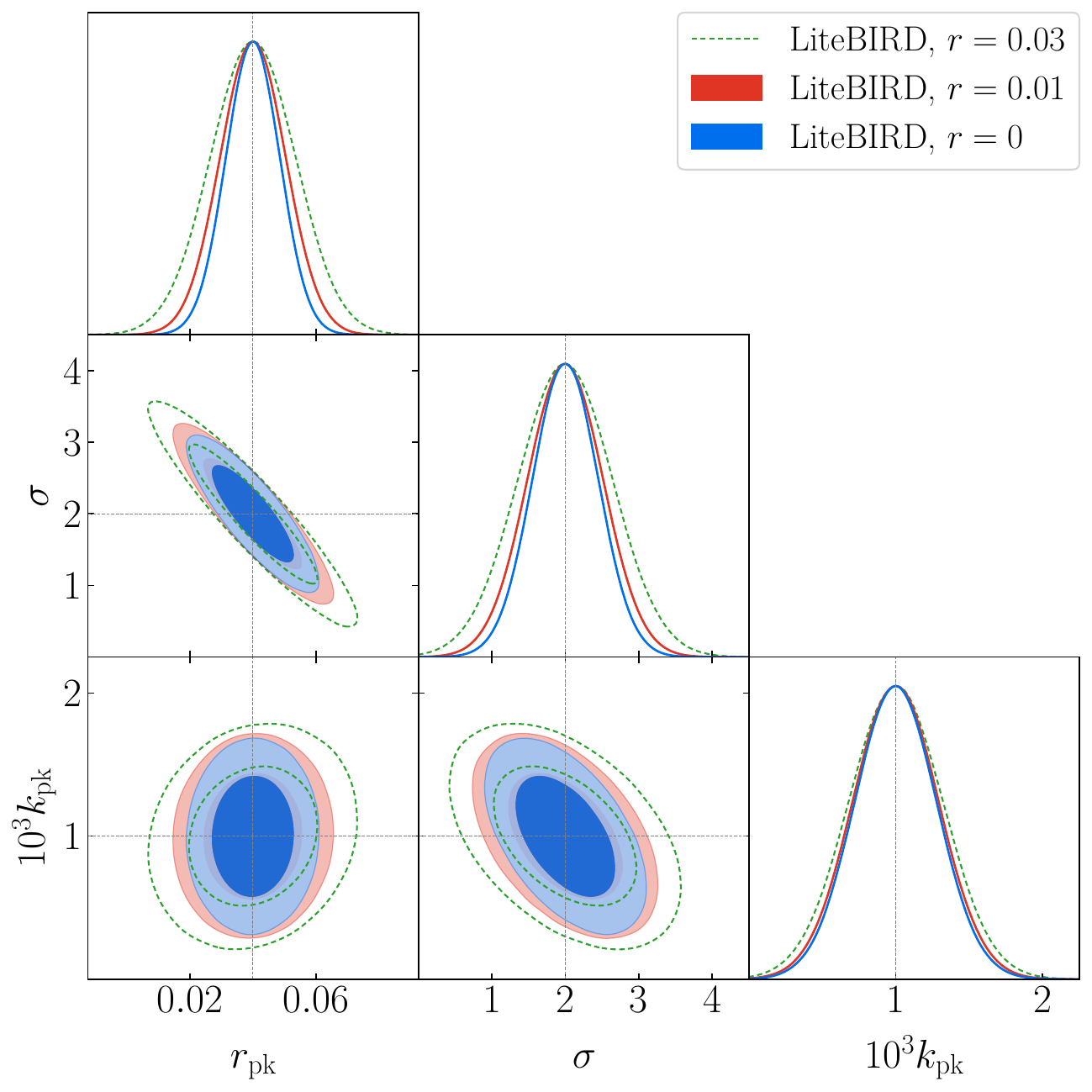}
    \includegraphics[width=0.46\linewidth]{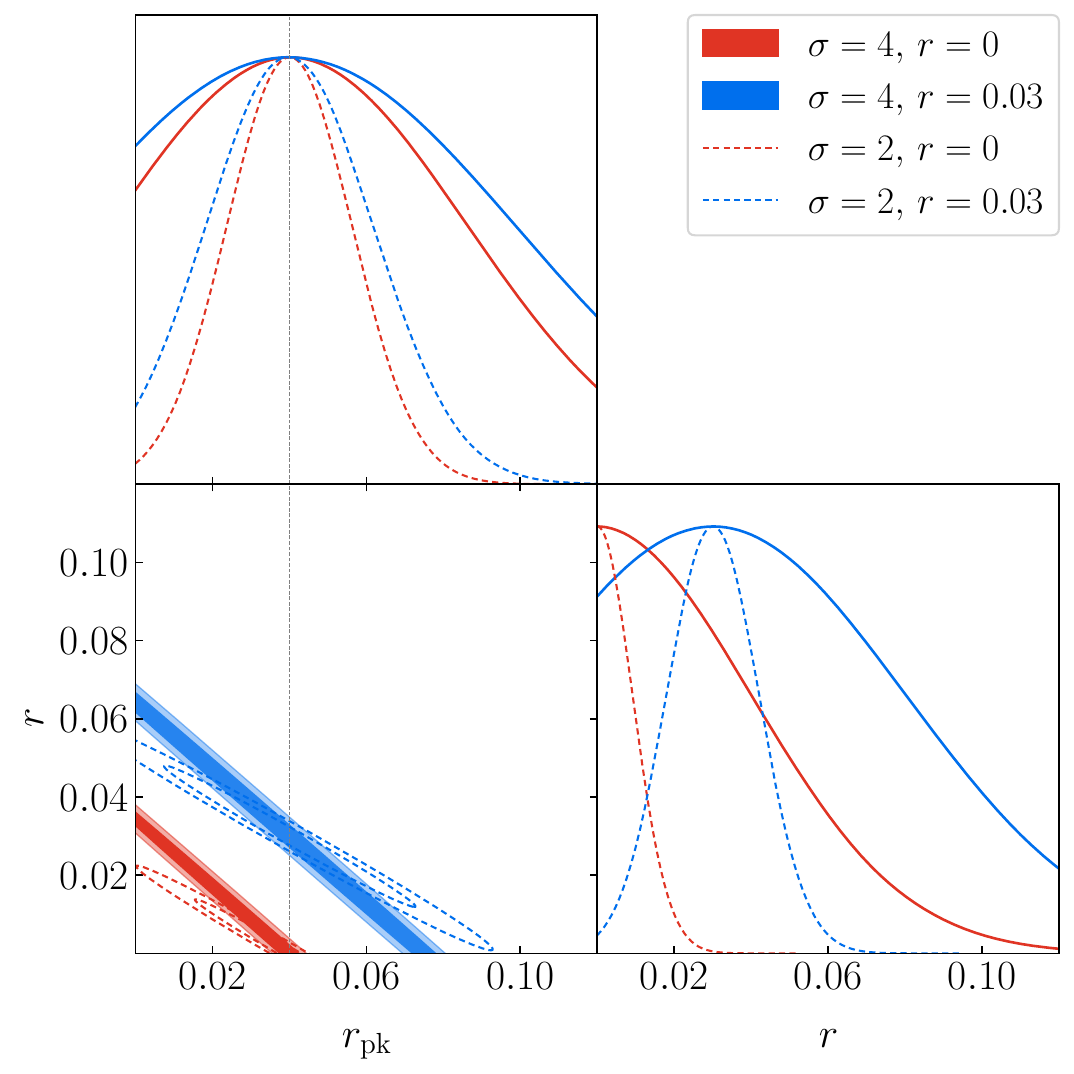}
    \includegraphics[width=0.46\linewidth]{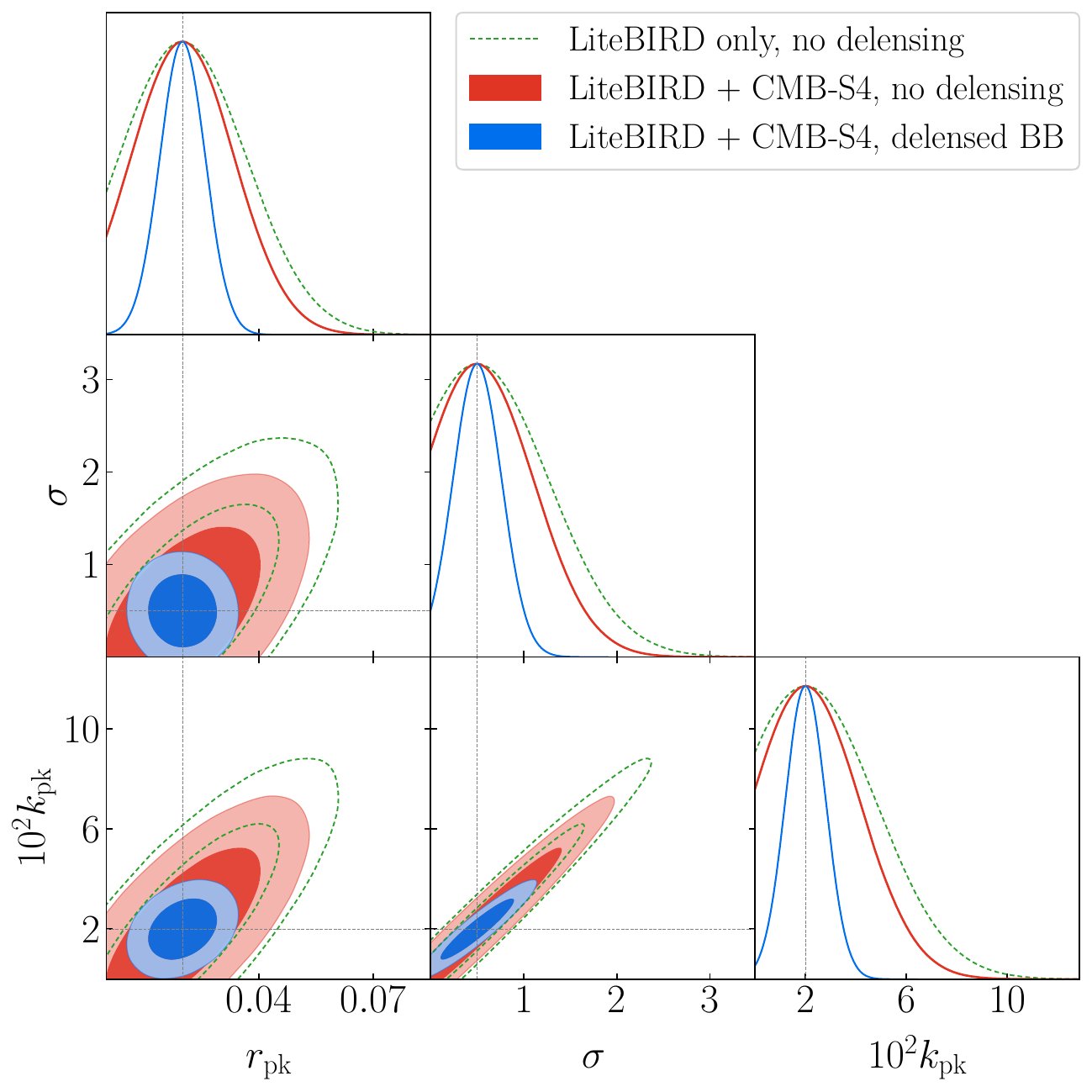}
    \caption{Left: Fisher forecast for the tensor parameters taking $\rp=0.04$, $\sigma=2$, $k_{\rm pk}=10^{-3}$ for three different values of the vacuum tensor scalar ratio $r$. No delensing assumed for this case. Right: Forecasts for $r$-$\rp$ with $\sigma$ and $\kpk$ fixed to the same values as in the left panel. Bottom: $\rp=0.02,\,\sigma=0.5,\,k_{\rm pk}=2\times 10^{-2}$ assuming $A_\mathrm{delens}=0.8$, $r=0$ for this case.}
    \label{fig:Fisher_sfsr_delens}
\end{figure}
The left panel of Figure~\ref{fig:Fisher_sfsr_delens} presents forecasts assuming the presence of the vacuum contribution to the tensor power spectrum with $r=0.03,\,0.01$ which effectively acts as a `noise' for the log-normal tensor spectrum detection. For the $r=0.01$ scenario, the forecasts are not very different from the case of $r=0$ which is expected since in this case the sourced tensor spectrum is much larger than the vacuum contribution on the relevant scales. {As expected, for larger $r$, the sensitivity to the log-normal tensor spectrum parameters is slightly reduced. The right panel of the same figure shows how the degeneracy between the parameters $r$ and $\rp$ increases as the peak width $\sigma$ increases, which leads to significantly worse estimates of $\rp$. This is not unexpected, since for larger $\sigma$ the lognormal spectrum starts to resemble a flat spectrum on the relevant scales.}

Finally, the bottom panel presents a scenario with a smaller amplitude of the sourced tensors $\rp=0.02$, $\sigma=0.5$ and $\kpk = 2\times 10^{-2}$. In this case the signal is detectable only if a significant amount of delensing can be achieved. For this value of $\rp$, the $B$-mode signal is entirely lensing dominated on angular scales $\ell>100$.

\section{Conclusions \label{sec:conclusion}}
In the event of a future $B$-mode detection associated to primordial tensor perturbations, the natural next step would be to understand the production mechanism behind these perturbations. This information is contained in the amplitude, shape, chirality and non-Gaussianity of the primordial tensor spectrum. 

Our focus here has been on the shape of the tensor spectrum, taking the example of a bump-like feature typical of GW sourced by axion-gauge fields during inflation. This shape deviates sharply from the SFSR prediction which is bound by the tensor consistency relation to be a power law with a slightly red tilted spectrum. 

Naturally, a detection of such a feature in the primordial tensor spectrum would hint to inflationary dynamics richer than those of SFSR and could be used to constrain the parameter space of such axion-gauge field models. On the other hand, even the non-detection of such a spectrum would be quite informative since that would strengthen our confidence in the SFSR model, especially given the difficulty in directly verifying the SFSR tensor consistency relation with future CMB probes. 

In this paper, we first searched for the presence of such features using the temperature and polarisation anisotropy data from the \textit{Planck} + BICEP/Keck experiments. While the current data do not provide any evidence for the presence of such features, they do place constraints on the tensor power spectrum parameters. In particular, strong constraints on the peak amplitude are obtained in the region $10^{-3}\lesssim k \lesssim 10^{-2}$ which is main sensitivity range of the BICEP/Keck $B$-mode data. 

We also presented forecasts of the ability of two future CMB experiments, namely LiteBIRD and CMB-S4, to detect such features. LiteBIRD's unprecedented accuracy in measuring polarisation at large angular scales gives it an excellent sensitivity to the amplitude of such features and will improve upon current sensitivity by at least a factor of two. However, accurate measurements of the width will require a significant amount of delensing, which could be provided by high-resolution CMB experiments such as CMB-S4. 

If primordial tensor power spectrum features were realised in Nature, their detection would represent a fascinating window into the earliest moments of the Universe, and help us get closer to a more complete understanding of the physics of inflation.  With the CMB experiments coming online in the next decade, we might just be able to take a peek.

\section*{Acknowledgements}
This research includes computations using the computational cluster Katana supported by Research Technology Services at UNSW Sydney~\cite{Katana}. AM would also like to thank Giovanni Pierobon for useful discussions regarding Katana.

\appendix

\section{\boldmath\texorpdfstring{$\Lambda$}{}CDM parameter estimates}
\begin{figure}[ht]
    \centering
    \includegraphics[width=0.8\linewidth]{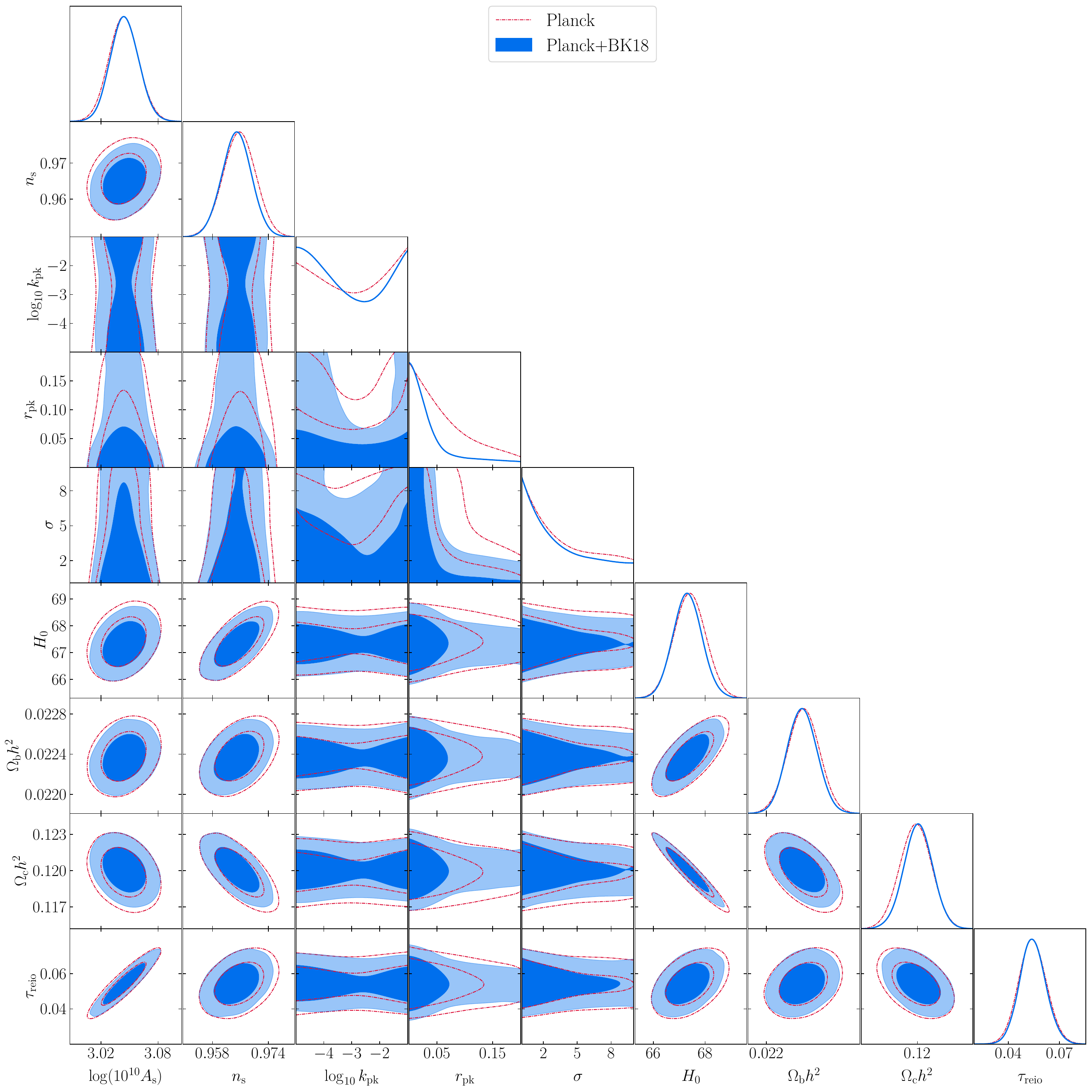}
    \caption{Marginalised $68\%$ and $95\%$ contours for $\Lambda$CDM and tensor spectrum parameters.}
    \label{fig:triangle_all}
\end{figure}

\begin{table}[ht]
\centering
\renewcommand\arraystretch{1.1}
\begin{tabular} { l  c}

 Parameter &  68\% limits\\
\hline
{\boldmath$\log(10^{10} A_\mathrm{s})$} & $3.045\pm 0.014   $\\

{\boldmath$n_\mathrm{s}   $} & $0.9649\pm 0.0040$\\

{\boldmath$H_0            $} & $67.3\pm 0.53        $\\

{\boldmath$\Omega_\mathrm{b} h^2$} & $0.02237\pm 0.00014$\\

{\boldmath$\Omega_\mathrm{c} h^2$} & $0.1201\pm 0.0012$\\

{\boldmath$\tau_\mathrm{reio}$} & $0.054\pm 0.0073 $\\
\hline
\end{tabular}
\caption{$\lcdm$ parameter means and $68\%$ limits for the log-normal model using \textit{Planck}+BK18 data.}
\end{table}

\bibliographystyle{JHEP}
\bibliography{ref}

\end{document}